\documentclass[11pt]{article} \usepackage[utf8]{inputenc}
\usepackage{amsmath,amsfonts,amssymb,latexsym} \usepackage[T1]{fontenc} 
\newtheorem{satz}{Theorem}[section] 
\newtheorem{assumption}[satz]{Assumption}

\newtheorem{defi}[satz]{Definition}

 \newtheorem{bem}[satz]{Remark}
\newtheorem{lemma}[satz]{Lemma}
 \newtheorem{koro}[satz]{Corollary}
\newtheorem{conclusion}[satz]{Conclusion}
 \newtheorem{ob}[satz]{Observation}

\newtheorem{conjecture}[satz]{Conjecture}

 \newcommand{\mbf}{\mathbf} 

\newcommand{\tit}{\textit}

 \newcommand{\R}{\mathbb{R}}
 \newcommand{\Z}{\mathbb{Z}}

\newcommand{\beq}{\begin{equation}}
 \newcommand{\eeq}{\end{equation}}
 \begin{document}
\thispagestyle{empty} \begin{center} \vspace*{1.0cm} {\Large{\bf Quantum Theory as emergent from an undulatory\\ translocal Sub-Quantum Level}} \vskip 1.5cm

{\large{\bf Manfred Requardt}}

\vskip 0.5cm

Institut fuer Theoretische Physik\\ Universitaet Goettingen\\ Friedrich-Hund-Platz 1\\ 37077
Goettingen \quad Germany\\ (E-mail: requardt@theorie.physik.uni-goettingen.de)

\end{center}

\begin{abstract}We argue that quantum theory is a low-energy effective theory which emerges from some sub-quantum level theory which is of an undulatory and translocal character. We show the close connection of quantum theory with both gravity and the holographic principle which are different phenomena of one and the same theory on this primordial level. An important role in our analysis is played by the concept of a generalized renormalization group connecting this primordial level and e.g. quantum theory plus a continuous space-time. We show that characteristic phenomena like the seemingly instantaneous state reduction, the EPR-paradox or the problem of polydimensions can be understood in our undulatory translocal theory in a realistic way. Most importantly, we give a realistic interpretation of the phasefunction as a collective action variable in the spirit of Bohm and explain the emergence of a macroscopic notion of time. 
\end{abstract}
 \newpage \setcounter{page}{1}
\section{Introduction}Originally, quantum theory was an amalgamation of, at first glance, two
quite distinct bundles of ideas, the different underlying concepts and ideas being well expressed in
the different denotations, i.e. \tit{quantum mechanics} (Bohr, Heisenberg etc.) and \tit{wave
mechanics} (D'Broglie, Schroedinger) with both being finally absorbed in the so-called \tit{Copenhagen
interpretation}. In the latter one, however, many of the original undulatory ideas have been more or
less eliminated as fundamental concepts. What remained is a very abstract framework which is openly
directed against the assumption of an underlying ontological model on some subquantum level which is able to
implement all the strange phenomena quantum theory is predicting in our macroscopic world. We think
everybody is aware of the hot dispute between Bohr and Schroedinger and the critical remarks of e.g.
Heisenberg in the formative period of quantum theory (cf. the beautiful biography of Schroedinger by
Moore, \cite{Moore}).

A cornerstone of the Copenhagen interpretation and the \tit{wave-particle dualism} is the observation that it
 seems to depend on the kind of experiment, being performed, whether the wave- or the particle-like properties 
of quantum matter are exhibited while
nothing can be known about the presumed quantum object as such. It is even sometimes denied that
such  carriers, having these seemingly contradictory properties, do actually exist. 
There does however not only exist a certain tension between the wave- and the particle-picture but there does exist in fact a more fundamental antagonism as to the appropriate structure of a well-formed fundamental theory of microphysics. We would like to call the two working philosophies the \tit{top-down approach} and the \tit{bottom-up approach}. The first paradigm was for most of the time presumably dominant in modern physics. It regards particles, waves and fields as essentially separate entities which moved through some background (i.e., space-time) which was considered as being something completely different. The latter world view, the one we prefer, regards the so-called \tit{quantum vacuum} as the ultimate primordial structure and considers what we like to call particles  and all the other higher structures as forming a hierarchy of increasingly complex excitation patterns emerging from this primary cause.
Such ideas can for example be already found in \cite{Wheeler}, section 44.3ff and go back to Clifford and Riemann, to menton only a few. A brief quote from \cite{Wheeler} p.1203 reads:
\begin{quote}
Particles do not form a really basic starting point for the description of nature. Instead they represent a first-order correction to vacuum physics\ldots.
\end{quote}

One should note that at the moment there seems to occur a certain back-swing. In the sequel of the \tit{holographic principle} a reinterpretation of gravity is emerging with certain consequences also for quantum theory. To mention some representative papers, see \cite{Adler},\cite{Pad},\cite{Verlinde}. There was even a recent conference with the title ``Emergent Quantum Mechanics'' (\cite{Vienna}). However, as far as we can see, most of the quantum literature, we are aware of, envokes a (stochastic) particle picture.

In contrast to that, our plan is to develop in the following such a unified model of quantum theory on some subquantum
level which is based on the undulatory picture advocated originally by Schroedinger (an earlier preliminary version is \cite{Requ7}; see below). Among other
things, we want to show that the so-called particle-like phenomena are rather attributes of an
underlying undulatory microcosmos. In this context we mention a bundle of ideas which were developed
by Bohm in his book \cite{Bohm} and which turn out to be surprisingly akin to our own findings. We
note, however, that there seem to exist in fact serious reasons why this undulatory picture has been largely
neglected in the process of canonization of the framework of quantum theory (see the following
remarks).
 \begin{ob}The arguments against a realistic interpretation of quantum theory and the
complex wave function are certainly convincing as long as the prevailing ontological framework is
not changed. This concerns in particular the nature of space-time on the microscopic scale.
 \end{ob}

The following quotations reflect the point of view Schroedinger had adopted and are from
\cite{Moore}: \begin{quote} \ldots The Broglie-Einstein wave theory of the moving particle,
according to which the latter is nothing more than a kind of whitecap on the wave radiation that
forms the basis of the world\ldots\\ \ldots The world is based on wave phenomena , while particles
are mere epiphenomena\ldots\\ \ldots He neglected the duality of wave and particle in favor of a
world of waves alone\ldots \end{quote} (see \cite{Moore} p.188, p.312).

As we already said, there do exist serious reasons against such a point of view. To mention a few, there
was the problem of \tit{polydimensions}, mentioned e.g. by Heisenberg (at the fifth Solvay Conference, October 1927, \cite{Moore} p.241):
\begin{quote}\ldots I see nothing in the calculations of Mr. Schroedinger that justifies his hope
that it will be possible to explain or understand in three dimensions the results from
polydimensions\ldots
 \end{quote}
On the other hand, Schroedinger said at the same conference
\begin{quote}\ldots each individual can be distributed through all space in such a way that the individuals interpenetrate\ldots
\end{quote}
We will develop such ideas in the main part of our paper.

 Another important problem was the so-called apparent \tit{instantaneous breakdown of the wave packet}
 in the measuring process or the particle-like behavior in the detection process of a quantum object in e.g.
 a photographic plate or in a Wilson chamber (more on these topics in the following).

On the other hand, as an example of a completely different point of view we mention the beautiful
discussion about the reality of wave functions in \cite{Weinberg}, p.61ff, we quote Scrooge from
p.63:
 \begin{quote}
\ldots It seems to me that none of this forces us to stop thinking of the wave
function as real; it just behaves in ways that we are not used to, including instantaneous changes
affecting the wave function of the whole universe\ldots 
\end{quote}
 Weinberg says he is very
sympathetic with this point of view as we are too. On should for example note that it is an (in our view) illusion that in the course of a position measurement a particle is observed at, say, position $x$. One does in fact really never observe a particle but rather a sharply peaked new wave function. Typical position measurements are of the order of some microns. That is:
\begin{conjecture}The ontological building blocks of quantum theory are the wave functions. In the course of a measurement we observe the (almost instantaneous) transition of some wave function into a transformed or reduced new wave function.
\end{conjecture}

 Furthermore, it becomes more and more evident in
e.g. quantum field theory in curved space-time that the particle picture is of a quite limited value
and is certainly not fundamental (cf. \cite{Wald}). See also the discussion in \cite{Cao} about the
field and/or the particle ontology.

What we have said up to now clearly shows that if one wants to attribute a kind of reality to the
wave function or wave field one has to explain within an undulatory picture phenomena like
instantaneous state reduction, particle-like behavior in general and e.g. the problem of
polydimensions. In this connection we want to point to a phenomenon which is in our view not
sufficiently appreciated (while it intrigued Schroedinger in the early days of quantum theory).
\begin{ob}It is in our view remarkable that energy and momentum are always intimately related to
frequency and wave number of an undulation pattern. This is almost impossible to understand within a
particle philosophy and reminds us of the apparent coincidence of proportionality of heavy and
inertial mass in classical mechanics before the advent of general relativity.
\end{ob}
 See e.g. Schroedinger in a letter to Wien
(\cite{Moore} p.226):
 \begin{quote} \ldots I have the feeling -- to express it quite generally --
that we have not yet sufficiently understood the identity between energy and frequency in
microscopic processes\ldots \end{quote}

In the following analysis the role of the \tit{phase function}, $S(x_1,\ldots,x_n;t)$, and the
complex structure of quantum theory in general is of central importance. As to this topic the
statement of Dirac in \cite{Dirac} is remarkable. In this paper Dirac is concerned with what is
actually \tit{the} characteristic feature of quantum theory (typically the non-commutativity of
observables is considered as the most remarkable phenomenon). Dirac however has a different opinion
and says \begin{quote}\ldots this phase quantity, which is very well hidden in nature and it is
because it was so well hidden that people had not thought of quantum mechanics much earlier\ldots
\end{quote} See also Yang in \cite{Yang} and the interesting discussion between Ehrenfest and Pauli
(\cite{Ehrenfest},\cite{Pauli1},\cite{Pauli2}).

We conclude this introductory section with a remark: \begin{bem}There is a widespread (slightly
arrogant) attitude concerning notably the philosophy of e.g. Schroedinger, classifying it as a
little bit old-fashioned ( the same holds for example for Einsteins point of view concerning quantum
theory). We think, if their utterances are really read carefully one must come to a different
oppinion. \end{bem} (Dirac loc.cit.): \begin{quote} \ldots One should not build up ones whole
philosophy as though this present quantum mechanics were the last word\ldots \end{quote}
 \section{Bohm's Wholeness and the Implicate Order}
 In the next sections we will develop a coherent framework based on an undulatory picture underlying quantum theory as we
know it. On some subquantum level it will comprise among other things a reformulation of microscopic
space-time as kind of a wormhole space, i.e. with built in translocal behavior, implementing the
\tit{holographic principle} and employing ideas developed in \cite{Sync} and \cite{Watts} concerning
translocal networks of so-called \tit{phase oscillators}. But before we will embark on this
enterprise, we would like to mention a couple of ideas which have been formulated by Bohm in his
book \cite{Bohm} because they support our undulatory philosophy.
 \begin{bem}We want to emphasize the
point that we do not have in mind the so-called Bohmian mechanics which is in a sense a model based
on point particles guided by some wave field. Bohm himself made some remarks as to this model in
\cite{Bohm} to which we refer the reader. We will rather concentrate on his ideas about a possible
underlying undulatory picture.
 \end{bem}

Ideas which are closely related to our own point of view are developed in \cite{Bohm} beginning with
section 4.9 (p. 87 ff). Bohm introduces an underlying substrate on some subquantum level consisting
of a nested hierarchy of vacuum fluctuations with fluctuation patterns on a certain scale in a
certain region of space and time being averages over the fluctuations of the next lower level while
the patterns we usually call 'particles' being relatively stable excitations on top of this sea of
vacuum fluctuations. He then introduces the idea that on each level of \tit{coarse graining} the
excitations on that scale of resolution of space-time determine their own motion largely independently
of the degrees of freedom (DoF) on the finer scales of resolution (i.e. we have what one would call
a hierarchy of scale dependent \tit{effective theories}). We developed such a model theory in quite
some detail in \cite{Requ1}, calling it a \tit{geometric renormalization group}, see also
\cite{Requ2}.

In a similar vein he introduces the concept of \tit{collective averaged coordinates} in
\tit{(quantum) space-time}, describing the large-scale behavior and being made up of the DoF of the
more microscopic levels. These ideas are inspired by earlier work of him and Pines in many-body
theory (\cite{Pines}). \begin{quote} \ldots Thus, the presence of matter as we know it on a large
scale would mean the concentration of a non-fluctuating part of the energy\ldots on the top of the
infinite zero-point fluctuations of the 'vacuum field' \ldots \end{quote} He then proceeds,
beginning with p.96, to provide an interpretation of the \tit{phase function}, $S$,
($\psi:=Re^{iS/\hbar}$). \begin{quote} \ldots Let us suppose\ldots that in each region of space and
time associated with any given level of size there is taking place a \tit{periodic inner
process}\ldots This periodic process would determine a kind of \tit{inner time} for each region of
space, and it would therefore effectively constitute a kind of \tit{local clock}\ldots \end{quote}

On p.97 he mentions the famous example of two clocks or oscillators coming in phase (an observation
once made by Huyghens, cf. \cite{Sync} p.106 or \cite{Watts} p.223, and speaks on p.99 of the \tit{infinity of clocks
within clocks} (an idea which is very simlar to our idea of a geometric renormalization group,
\cite{Requ1}) and of the phase function, $S$, as a function of this clock structure. Finally,
starting with section 6.9, he introduces the idea of a \tit{hologram} and the concept of
\tit{implicate order} and the \tit{holomovement}.
 \section{The Substrate of Vacuum Fluctuations and its Microscopic Dynamics}
 We saw that in Bohm's framework the hierarchy of
vacuum fluctuations plays a central role as the basis from which all the more mesoscopic or
macroscopic patterns (relative to some primordial \tit{Planck scale}) do emerge. In a similar vein we
introduced and employed this quantum substrate in section 2 of \cite{Requ3}. We think
the microscopic hierarchical structure of vacuum fluctuations plays a role in various areas of
modern high energy physics, from the \tit{cosmological constant problem} to the mysteries of
\tit{quantum entanglement} and \tit{holography} in quantum gravity. Some of these questions we
addressed in \cite{Requ3}.

The crucial point in our view is the following observation (which is to our knowledge not really
appreciated or rather overlooked)
 \begin{ob}In a realistic interpretation the pattern of vacuum fluctuations is a self-exciting medium which  is strongly
(anti)correlated on all scales. This means, it exhibits, according to our analysis, a very rigid
long-range order of positive and negative (e.g. energy) fluctuations on all scales which more or
less compensate each other.
 \end{ob}
 \begin{koro}Ths is, in our view, the reason that for example
the cosmological constant is so small, in contrast to the naive estimates which, typically, simply
add up the individual contributions (we discussed this point in \cite{Requ4} and \cite{Requ5}.
\end{koro}
\begin{bem}Some papers discussing the structure of the quantum vacuum and/or the
cosmological constant problem are
\cite{Rugh1},\cite{Rugh2},\cite{Boyer},\cite{Weinberg2},\cite{Strau}.
 \end{bem}

The above observation is a result of a detailed technical analysis of fluctuation theory done in a
more general setting (cf. \cite{Requ4}). We begin with a simple model assumption. Assume we are
given a volume, $V$, in $\R^3$. Using Planck-units we have roughly $N_V:=(V/l_p^3)$ grains of Planck
size in $V$. Assuming, furthermore, that some quantity $q_i$ is given in each of the grains lablled
by $i$ with
 \begin{equation}<q_i>=0\quad \text{and}\quad <q_i^2>\quad \text{large but finite} \end{equation}
\begin{lemma}Assuming that the individual quantities, $q_i$, fluctuate almost independently, we have
(a consequence of the central limt theorem):
 \begin{equation}<Q_VQ_V>^{1/2}\sim (V/l_p^3)^{1/2}=N_V^{1/2} \end{equation}
 With $Q_V:=\sum_i q_i$. I.e., with $N_V$ gigantic for a
macroscopic volume, $N_V^{1/2}$ is still macroscopic. \end{lemma}
 \begin{ob}Such a behavior is not
observed for quantities like e.g. vacuum energy and the like. Quite to the contrary, such integrated
quantum fluctuations of physical quantities in the vacuum are typically of microscopic size in macroscopic volumes.
\end{ob}
 \begin{bem}Note that we made the assumption that we have microscopic DoF which are located
in grains of Planck size. This is an assumption in the line of our realistic working philosophy (in
contrast to the Copenhagen framework). \end{bem}

The following result corroborates these observations.
 \begin{ob}In quantum field theory the vacuum
state is an eigenvector of the Hamiltonian with $H\Omega=0$. This implies that not only
$<\Omega,H\Omega>=0$ but also $<\Omega,H^2\Omega>=0$, i.e. the total fluctuation exactly vanishes.
In our realistic interpretation this means that in every observation the sum of negative and
positive energy fluctuations exactly cancel each other. \end{ob}
 In \cite{Requ3}, conclusion 2.8 we derived a useful
a priori estimate concerning the expected amount of negative fluctuations in the vacuum.
\begin{bem}Note that in many heuristic discussions the vacuum fluctuations are treated in a very
missleading way. Frequently they are represented as virtual particle pairs emerging from the vacuum
with essentially positive energies being implied.
\end{bem}
 \begin{conjecture}We conjecture that the (necessarily existing)
negative contributions of the vacuum fluctuations are essentially of gravitational type. We will
discuss this in more detail elsewhere.
 \end{conjecture}
\begin{bem}An idea in this direction can e.g. be found in \cite{Wheeler2}, p.269.
\end{bem}

We can weaken the assumption of the dynamical independence of the local DoF considerably by assuming
that the fluctuations in the grains are not strictly independent but short-range correlated like
e.g.
 \begin{equation}|<q(x)q(y)>|\;\text{integrable}\; in\; |x-y| \eeq
 \begin{lemma}With $q(x)$ the
operator density of some physical observable and $Q_V:=\int q(x)\,d^nx$ the integral over $V$; if we
want to have that
 \beq <Q_VQ_V>^{1/2} \ll V^{1/2} \eeq
 it is necessary that
 \beq \int_V d^ny<q(x)q(y)>\approx 0 \eeq
for large $V$ (cf. \cite{Requ4}) \end{lemma}
 \begin{bem}For the numerical estimate it
is useful to perform the integration over a volume $V$ with a smooth boundary (see \cite{Requ4} or
\cite{Requ6}. \end{bem}
 \begin{conclusion}In order that the integrated fluctuations in a macroscopic
region, $V$, are negligible, the individual fluctuations on the microscopic scale have to compensate
each other quite effectively (i.e. spatially oscillating from positive to negative values and vice
versa). This holds in particular for the so-called vacuum fluctuations.
 \end{conclusion}
\begin{ob}The degree of vanishing of $\int_V d^ny<q(x)q(y)>$ with $V\rightarrow\R^n$ is encoded in
the degree of vanishing of the Fourier transform (FT) of $<q(x)q(y)>$, $\hat{F}(\mbf{k})$, near
$\mbf{k}=0$ (\cite{Requ4}).
 \end{ob}
 With
 \beq f_R(x):=f(|x|/R)\; , \; x\in \R^n \eeq
 $f\geq 0\; ,\;f=1\;\text{for}\; |x|\leq 1$, $f$ smooth and of compact support, we derived the following result in
\cite{Requ4}:
 \beq <Q_RQ_R>\,\sim R^n\cdot R^{-\alpha}\cdot\int
|\mbf{k}|^{\alpha}|\hat{f}(\mbf{k})|^2d^nk\quad\text{for}\; R\rightarrow\infty \eeq
 $\hat{f}$ the F.T.
of $f$ and
 \beq \hat{F}(\mbf{k})\sim |\mbf{k}|^{\alpha}\quad\text{near}\quad k=0 \eeq
 \begin{ob}We
see that the almost vanishing of fluctuations in macroscopic domains (compared to the Planck scale!)
can be achieved for correlations, $<q(x)q(y>$, provided the F.T. vanishes sufficiently fast near
$k=0$. Note however that this can hold for correlation functions which are integrable (short-range
decay).
 \end{ob}

One can considerably strengthen this result by employing the \tit{holographic principle} which
roughly says in some version that the number of available DoF in a volume, $V$, is only proportional
to the area of the boundary of $V$. We analyzed its implications in some detail in \cite{Requ3} and
\cite{Requ4} with the result: \begin{conclusion}The above observation of the almost vanishing of
vacuum fluctuations in macroscopic regions together with the implications of the holographic
principle lead inevitably to the following result: The vacuum fluctuations are both strongly
(anti)correlated and display a translocal long-range order of a peculiar geometric type.
\end{conclusion}

This \tit{translocal} long-range order will play an important role in the following. So it is
appropriate to say some more words about it. We refer to \cite{Requ3},\cite{Requ4} concerning
technical details. What picture we have in mind can perhaps best be expressed by the following
quotation (\cite{Witt}).
 \begin{quote}
\ldots But if a wormhole can fluctuate out of existence when
its entrances are far apart \ldots then, by the principle of microscopic reversibility, the
fluctuation \tit{into} existence of a wormhole having widely separated entrances ought to occur
equally readily. This means that every region of space must, through the quantum principle, be
potentially ``close'' to every other region, something that is certainly not obvious from the operator
field equations which, like their classical counterparts, are strictly local.\ldots It is difficult
to imagine any way in which widely separated regions of space can be ``potentially close'' to each
other unless space-time itself is embedded in a convoluted way in a higher-dimensional manifold.
Additionally, a dynamical agency in that higher-dimensional manifold must exist which can transmit a
sense of that closeness
 \end{quote}
 \begin{defi}We call such a space-time structure a wormhole
space. I.e., below the smooth surface of macroscopic behavior the microscopic space-time structure
is assumed to have both a near-order among the individual space-time grains (implying some local
interaction among the  DoF in the infinitesimal neighborhood of each grain) and a far-order
accomplished by a certain fraction of elementary interactions extending to other regions of
space-time which are some distance away measured in macroscopic units.
 \end{defi}
\begin{bem}We want to stress the point that what we have in mind are microscopic wormholes or, more specifically, microscopic structures which act as wormholes.
\end{bem}

Within the framework of \tit{structurally dynamic (random) networks} we studied such erratic systems
already in \cite{Requ1} and \cite{Requ2}, in which we generalized the more determinstic and regular
cellular automata models , introduced by 't Hooft (see e.g. \cite{Hooft1},\cite{Hooft2} and further
literature mentioned there). In \cite{Now} we made a large-scale numerical analysis of the dynamical
and statistical behavior of such large networks. It is an interesting coincidence that roughly at
the same time such translocal network architectures (usually of a static type, i.e., with a fixed
wiring) have been studied in network theory, called \tit{small world networks} (cf. \cite{Sync} and
\cite{Watts}, see also the literature cited in \cite{Loch}).
\section{The Translocal Network of Clock-Like Phase Oscillators as a Basis of Space-Time}
In \cite{Requ1},\cite{Requ2},\cite{Now} and further literature cited there we studied various discrete dynamical laws governing the evolution of our statistical networks which already displayed some \tit{undulatory} behavior. When we analysed the dynamics of these network laws, we realized that they are structurally similar to large networks of coupled oscillators and this is the route we will persue in the following. Crucial is in our analysis the idea of a \tit{geometric renormalization group}. In the above papers we developed an algorithm which allows to make systematic coarse-graining steps in the network architecture. That is, we construct a hierarchy of increasingly coarser and, at the same time, smoother levels in the original network which represent, on the other hand, new networks. This is achieved by a method of \tit{averaging}.
\begin{ob}In this process of averaging at each step certain clusters of nodes and bonds, called cliques by us, are grouped together to yield the local neighborhoods of the next level. On the other hand, there always remain bonds which do not belong to these local neighborhoods but connect neighborhoods which are some distance apart. this then leads finally to the (hidden) non-local structure of quantum theory after several steps of such coarse-graining.
\end{ob}
\begin{defi}i) A clique is geometrically a complete subgraph, i.e. a graph with each pair of nodes connected by an edge.\\
ii) A clique graph is a coarse-grained graph in which the nodes are the cliques of the preceding graph and the edges (bonds) are given by the overlap of cliques. In our approach we defined edges via a sufficient (non-marginal) amount of overlap (for details see our cited papers). 
\end{defi}
Clique graphs, $C(G)$, with $G$ a graph and their hierarchy, $C^n(G)$, are an interesting field of research in graph theory.
\begin{ob}In \cite{Requ1} we showed that these hierarchies may have fixed points! In our philosophy the possible fixed points of $C^n(G)$ over appropriately chosen graphs $G$ are assumed to describe potential smooth macroscopic space-times.
\end{ob}
The above concepts remain of course for ever within the regime of networks or graphs. In order to get something continuous another step is necessary. This program we formulated and described in detail in \cite{Requ8}.
\begin{ob}In order to arrive finally at something which represents a continuous space, one has to augment our renormalization steps with an algorithm which allows to perform a geometric continuum limit. We accomplished this in the above cited paper by using coarse geometry and ideas of Gromov.
\end{ob}   

Expressed in a more physical language, in this process the node- and bond-states of the coarser networks become kind of \tit{collective DoF} formed from the DoF of the next lower network, just in the way as envisaged by Bohm. The philosophy of the renormalization group is realized in the consecutive coarse-graining steps which average over short-range fluctuations in the network, viz., the short range vacuum fluctuations and leave us in the end with a scale of resolution which is assumed to correspond to the typical scale on which the processes of ordinary quantum theory as we know it  take place.
\begin{bem}In this process, as in the usual renormalization group formalism, the dynamical (network) laws are also renormalized, viz., are scale-dependent.
\end{bem}   
If we transform our cellular network into an array of interacting oscillators we observe the following. We can relate suitable infinitesimal lumps of nodes and bonds (for example the clique-subgraphs), dubbed $C_i$, to elementary oscillators, while we relate the strength of interaction among these lumps to the degree of interaction between the corresponding clusters of oscillators. As these lumps are structurally similar but not completely identical (as a consequence of the element of randomness in the local structure and dynamics of our networks), we expect that the natural frequencies, $\omega_i$ (see below) of these oscillators scatter around some mean-value, $\bar{\omega}$, i.e. we will have
\beq \omega_i=\bar{\omega}+\Delta\omega_i   \eeq  
In the vicinity of the Planck scale the interaction among the oscillators is supposed to be strong and quite erratic. That is, the influence of the environment (given by the other oscillators) on an individual oscillator will deform this natural frequency appreciably and, furthermore, will strongly fluctuate in time. After several coarse-graining steps these collective effects will become both weaker and smoother so that the assumptions we will make below seem to be justified.

 Large networks of coupled oscillators have been studied  in general network theory not aimed at simulating the behavior of the \tit{quantum vacuum}. Such networks of \tit{phase oscillators} occur frequently in various fields of e.g. biophysics and are described in \cite{Watts} and \cite{Sync}, see also \cite{Stro1}.

The notion  \tit{phase oscillator} means, that the individual oscillators of a network are damped in such a nonlinear way that their amplitude happens to be energetically locked or, more generally, are confined to a \tit{limit cycle} after some \tit{transient} period  by a certain mechanism so that an individual oscillator has finally only its phase as DoF with (in two dimension and for the simplest case)
\beq d\theta_{ij}/dt=\omega_{ij}    \eeq
($\theta_{ij}$ the phases, $\omega_{ij}$ the frequencies, $(ij)\in \Z^2$)
The non-trivialty of the network model arises from the interaction between the oscillators in the network. A typical example is the \tit{Kuramoto model} (\cite{Stro1}, the true Kuramoto model is rather of \tit{mean field type}).
\beq  d\theta_{ij}/dt=\omega_{ij}-\alpha\cdot\sum_{kl}J_{ij,kl}\cdot \sin(\theta_{ij}-\theta_{kl})   \eeq
A very nontrivial example of an individual oscillator having a \tit{limit cycle} is the so-called \tit{van der Pol oscillator}
\beq \ddot{x}-\varepsilon(1-x^2)\dot{x}=0  \eeq
which for every $\epsilon>0$ has a stable limit cycle in the \tit{phase plane} of $(x,\dot{x})$ or a periodic $x(t)$ (cf. \cite{Stro2} or \cite{Jackson}). For small $\varepsilon$ the limit cycle is nearly circular and the frequency nearly a constant. In general the oscillation is however not sinusodial and the solution curve quite complicated (see below concerning the so-called \tit{averaging method} which produces a good approximation for small $\varepsilon$). 

Most of the interaction laws being studied are (in the tradition of ordinary statistical mechanics) of local (or nearest neighbor) type with $J_{ij,kl}$ only non-vanishing for $(ij),(kl)$ a small distance apart. What is interesting is the correlation between the oscillator phases
\beq C(r,t)=<\cos(\theta_{ij}(t)-\theta_{kl}(t)>  \eeq
with $r:=((i-k)^2+(j-l)^2)^{1/2}$, the averaging is made over pairs of oscillators in the network. Another form of \tit{order parameter} is
\beq R\cdot e^{i\phi}:=N^{-1}\cdot\sum_{j=1}^N e^{i\theta_j}  \eeq
($j$ running over the network sites).
\begin{bem}Note the formal similarity with the complex wave function in quantum mechanics.
\end{bem}  
We will not go into the details of the tedious numerical calculations of the various models (see, in particular, the paper by Niebur et al, \cite{Nie}). What is important is that an addition of a certain amount of long-distance wiring (interactions, connecting points in $\Z^2$, which are a large distance apart) leads to a stable \tit{phase-locking} among the oscillators over large distances and times, i.e., to a \tit{synchronization} of the individual phases. This is in marked contrast to models in which the translocal wiring is absent.
\begin{ob}It is important that our geometric renormalization group algorithm generates quasi automatically such a dual structure of, on the one hand, a network of densely entangled lumps which yields the local neighborhood structure plus a superimposed sparse network of translocal connections. Thus we exactly get a network type of the above structure which is also called a small-world network (cf. \cite{Loch}).
Furthermore, our networks are of the structurally dynamic type, i.e., also the wiring is (clock-)time dependent.
\end{ob}

 The network types we described above have an obvious similarity to the network of elemetary clocks, envisaged by Bohm in \cite{Bohm} with the clocks being essentially phase-oscillators.
\begin{ob}There exists however an important difference. The crucial ingredient of translocal connections seems to be absent in Bohms approach. Our above discussion shows that a certain amount of translocal wiring is essential for the coherence and synchronization of such a network of local elementary 'clocks'.
\end{ob}
\begin{conjecture}We conjecture that such a translocal wiring is the basis of the emergence of a smooth and coherent space-time manifold on the macroscopic scale, a structure we dubbed wormhole space.
\end{conjecture} 

The above Kuramoto network model of phase oscillators is an idealization in so far as only the oscillator phases are dynamic DoF. We expect that in our simulation of the fluctuating quantum vacuum by a large array of oscillators the interaction among the oscillators is not so weak that only the phases are affected. We rather think the following ansatz is more appropriate:
\begin{assumption}\label{undu}On a primordial network level where there does not yet exist a clear distinction between a certain near-order structure (typical of a smooth continuous space) and the translocal sparse contributions, or, more generally, as long as we do not regard the disordered network as being embedded in a smooth continuous space-time, we make the following ansatz for the undulation pattern at oscillator site $(i)$.
\beq \text{Amplitude:}\quad A_i(t)=A_0+\delta A_i(t)+a_i(t)    \eeq
\beq \text{Phase:}\quad \phi_i(t)=\omega_0 t+ \varphi_i(t)+\theta_i(t)    \eeq
\beq q_i(t)=A_i(t)\sin\phi_i(t)   \eeq
for some quantity $q_i(t)$ the nature of which is not important at the moment (cf. the remark in \cite{Bohm} on p.96).

\end{assumption}
\begin{bem}The intricacies which emerge if we embed such a structure in a continuous background are discussed below.
\end{bem}
The meaning of the above formulas is the following: $A_0,\omega_0$ are some mean values of Planck size. $\delta A_i(t),\varphi_i(t)$ are position dependent deviations from these average values, $A_0,\omega_0 t$, which have also Planck character in so far as they vary on a very short (Planck) scale in the network, i.e.
\beq A_0\gg |\delta A_i(t)|\quad ,\quad \omega_0t\gg |\varphi_i(t)|  \eeq
but
\beq |\delta A_i(t)-\delta A_j(t)|\gg |a_i(t)-a_j(t)|\quad ,\quad | \varphi_i(t)-\varphi_j(t)|\gg |\theta_i(t)-\theta_i(t)|  \eeq
\beq |\delta\dot{ A}_i(t)|\gg |\dot{a}_i|\quad ,\quad |\dot{\varphi}_i|\gg |\dot{\theta}_i|   \eeq
and $|a_i(t)|,|\theta(t)|$ are small compared to $A_0,\omega_0 t$. To put it briefly, $\delta A_i(t),\varphi(t)$ are still rapidly fluctuating on Planck scale while $a_i(t),\theta_i(t)$ are some slowly varying large scale modulations of the underlying rapid and short scale vacuum fluctuations.
\begin{ob}i) The superimposed large scale contributions, if they are present, are supposed to represent, after some appropriate coarse-graining (or renormalization) steps, the excitations we observe on, say, the scale of ordinary quantum theory. They have the character of amplitude plus phase modulation.\\
ii) If these patterns are absent we have the pure quantum vacuum, being described by the patterns given by $A_0,\omega_0,\delta A_i(t),\varphi_i(t)$.
\end{ob}  

We add a remark about an apparent circularity in the way we deal with the concept of time  in such models.  If we  introduce such  models  we must be able to talk about their dynamics, viz., we need some notion of (clock) time while talking  about the foundation of space-time itself. This is a constant problem in this field. We hence make the practical assumption that we either view the model as being embedded in some already existing space-time with a certain distance concept, or we apply the procedure we used in e.g. \cite{Requ1} and \cite{Requ2}. That is, we introduce a hierarchy of coarse grained levels in the primordial network and thus generate a macroscopic notion of space-time as an emergent structure via what we called a geometric renormalization group. We then can discuss the microscopic details of the network by referring to this coarse grained localization concept. This is in effect what one is essentially doing in quantum theory all the time.
 
As  $A_0,\omega_0$ are Planck scale values we make the reasonable assumption that
\beq \omega_0=\omega_{Pl}=E_{Pl}/\hbar   \eeq
with
\beq E_{Pl}\cdot t_{Pl}=\hbar\; ,\;p_{Pl}\cdot l_{Pl}=\hbar\; ,\;t_{Pl}=l_{Pl}/c\; ,\; l_{Pl}=(G\hbar/c^3)^{1/2}   \eeq

Such an undulation model was already introduced by us in \cite{Requ7} which, however, is rather of the type of a draft version. It contains a variety of (speculative) ideas and concepts concerning the emergence of quantum theory and, in particular, of the Schroedinger wave equation from an underlying substratum which was essentially modelled as a random network. However, various important points were only scetched in that paper. We were, in particular, unaware at that time of the vast field of oscillator networks and their fascinating properties. This unifying principle we want to provide in the following. First of all we want to show that the conjectured behavior of our local oscillators, i.e. small and slow fluctuations of amplitude and phase around a stable limit cycle representing a fast oscillation , is not really an extremely special assumption. It rather holds for a large class of examples of individual nonlinear oscillator exhibiting a self-sustained oscillation and is also the case for a network of such oscillators. 
\section{Two Time Scales and the Averaging Method for (Weakly) Nonlinear Oscillators}
The following is a mathematical interlude which serves as a justification of our undulating picture in such large networks of nonlinear oscillators. We want to show that an oscillating behavior for the individual oscillators of the type
\beq x(t)=(A_0+a(t))\cdot \sin(\omega_0 t +\theta(t))    \eeq
with small $a(t),\theta(t)$ (compared to $A_0,\omega_0 t$) is almost the generic case at least for a large class of oscillator models. 

We discuss the following class of oscillators:
\beq \ddot{x}+x=\varepsilon\cdot f(x,\dot{x};t)    \eeq
where $\varepsilon$ is assumed to be a small parameter. The technique of approximately solving such equations has a long and rich tradition. The problem is interesting because ordinary perturbation theory typically does not work because so-called \tit{secular terms} frequently do occur which grow with large $t$. Therefore other methods were invented. A classical source is \cite{Bog}. A very clear and concise account can be found in \cite{Jackson}, see also \cite{Stro2} where the (heuristic) method of \tit{two-timing} is discussed. The mathematical problems are treated in depth in e.g. \cite{Sanders} (with a brief history of the method in appendix A).

For $\varepsilon=0$ the general solution is 
\beq x=a\sin\phi\quad , \quad \dot{x}=a\cos\phi\quad ,\quad \phi:=t+\theta   \eeq  
We want to show that for $\varepsilon\neq 0$ the general solution can also be put in the form
\beq x=a(t)\sin\phi(t)\quad ,\quad \dot{x}=a(t)\cos\phi(t)  \eeq
We take the two-dimensional phase plane $\R^2$ with cartesian coordinates $(x,\dot{x})$. We define a map (polar coordinates)
\beq x=a\sin\phi\quad ,\quad \dot{x}=a\cos\phi  \eeq
(Note that at the moment $\dot{x}$ is not! the time-derivative of $x$).

We now take a solution curve, $x(t)$, of the original nonlinear equation and insert the $(a,\phi)$-parametrization. This then yields first-order differential equations for $(a(t),\phi(t))$.
\beq \dot{x}=\dot{a}\sin\phi+a\dot{\phi}\cos\phi=:a\cos\phi  \eeq 
\beq \ddot{x}=\dot{a}\cos\phi-a\dot{\phi}\sin\phi=-a\sin\phi+\varepsilon f(a\sin\phi),a\cos\phi;t)   \eeq
Multiplying the first equation by $\sin\phi$ and the second by $\cos\phi$ and adding yields 
\beq \dot{a}=\varepsilon f(a\sin\phi),a\cos\phi;t)\cos\phi  \eeq
\beq \dot{\phi}-1=\dot{\theta}=\varepsilon/a\cdot  f(a\sin\phi),a\cos\phi;t)ßsin\phi   \eeq
\begin{bem}Note that the differential equations for $a,\phi$ are first order as a consequence of our nonlinear coordinate transformation. Crucial in this respect is the term $a\cos\phi$ on the rhs of the first equation above.
\end{bem}

The important insight is now that if $\varepsilon$ is small the rhs of both equations are small so that both $a$ and $\theta$ are slowly varying.
\begin{ob}For  $\varepsilon$ small  both $a$ and $\theta$ are slowly varying.
\end{ob}
Here now enters the method of \tit{averaging}. With $\phi(t)=t+\theta$ we average over one period of $\phi$:
\beq (\bar{f}_c(a),\bar{f}_s(a)):=1/2\pi\int_0^{2\pi}d\phi f(a,\phi)\cdot(\cos\phi,\sin\phi)  \eeq
In lowest order in $\varepsilon$ we get:
\beq \dot{a}=\varepsilon/2\pi\int_0^{2\pi}f(a\sin\phi,a\cos\phi)\cos\phi\, d\phi  \eeq
\beq \dot{\phi}=1-\varepsilon/2\pi a\int_0^{2\pi}f(a\sin\phi,a\cos\phi)\sin\phi\, d\phi  \eeq
or
\beq \dot{a}=\varepsilon\cdot(\bar{f}_c(a)\quad , \quad  \dot{\phi}=1-\varepsilon/a\cdot \bar{f}_s(a)  \eeq
\begin{ob}Note that in lowest order both $\dot{a}$ and $\dot{\phi}$ depend only on $a(t)$ and not on $\phi(t)$.
\end{ob}

The general formulation by Krylov,Bogoliubov and Mitropolski consists in making a general ansatz:
\beq x=a\cos\psi+\varepsilon u_1(a,\psi)+\varepsilon^2 u_2(a,\psi)+\cdots    \eeq
with $u_k(a,\psi+2\pi)=u_k(a,\psi)$. For the so-called \tit{autonomous systems}, i.e.
\beq f(x,\dot{x};t)=f(x,\dot{x})   \eeq
they make an ansatz for $\dot{\psi}(t),\dot{a}(t)$:
\beq \dot{\psi}(t)=1+\varepsilon B_1(a)+\varepsilon^2B_2(a)+\cdots    \eeq
\beq \dot{a}(t)=1+\varepsilon A_1(a)+\varepsilon^2A_2(a)+\cdots  \eeq
retaining the feature that $(\dot{a},\dot{\psi})$ are only dependent on $a$.
\begin{bem}Such an ansatz applies for example for the (autonomous) van der Pol oacillator.
\end{bem}

More interesting in our case of an oscillator embedded in an array of other oscillators is the \tit{nonautonomous case}.
\begin{defi}The nonautonomous case is characterized by $\partial_t f(x,\dot{x};t)\neq 0$ .
\end{defi}
This case is more complicated but also more interesting and various problems can arise as e.g. \tit{resonance phenomena} and \tit{entrainment}. An example is
\beq \ddot{x}+x=\varepsilon f(x,\dot{x})+\varepsilon A\cos \Omega t    \eeq
which is of the type of the \tit{forced van der Pol equation}.
\begin{bem}New phenomena occur if the periodic force is not weak but strong. Strong forcing will generally lead to a less uniform behavior (cf. \cite{Jackson}). We remarked above (section 4) that we expect such an erratic behavior on the more primordial scales while the behavior is expected to become smoother and more predictable after several renormalization steps, viz., on the level where quantum theory applies.
\end{bem}
\begin{ob}[periodic forcing]In contrast to the simpler case of autonous systems, both $\dot{a}$ and $\dot{\psi}$ may now depend on $a(t)$ and $\psi(t)$.
\end{ob}
\begin{conclusion}With our original notation
\beq x(t)=(A_0+a(t))\cdot\sin(\omega_0+\theta(t))   \eeq
we hence have in the general case of an externally forced (viz., by the other members of the oscillator array) and self-exciting system that $a(t),\theta(t)$ are typically slowly varying in time and that the rhs of the differential equations 
\beq \dot{a}(t)=\cdots\quad , \quad \dot{\theta}=\cdots    \eeq
depend both on $a(t)$ and $\theta(t)$. We emphasize again that they are first order in time as a consequence of our nonlinear coordinate transformation, we made above.
\end{conclusion}
\section{Quantum Theory as an emergent Effective Mean Field Theory}
In this crucial section we are going to transform and unify the series of steps we have described in the preceding sections into a description of quantum theory as it appears to us on the smooth and coarse-grained scale we are accoustomed to. Central in this context is the transformation of the field of basic primordial quantities
\beq q_i(t)=A_i(t)\cdot \sin\phi_i(t)   \eeq
with $A_i(t),\phi_i(t)$ defined in assumption \ref{undu}
and the slowly varying large-scale contributions, $a_i(t),\theta_i(t)$, contained in the expressions for $A_i(t),\Phi_i(t)$, into quantities which are observable on the mesoscopic quantum scale. This transformation is highly non-trivial as these \tit{collective} quantities have to survive and are altered by the hierarchy of geometric renormalization steps which span many orders of magnitude.

To give an impression of what has to be achieved. In the primordial network, on which the $q_i(t)$ live, the wiring among the nodes $(i)$ is supposed to be very erratic and presumably rapidly changing with time. That is, nodes which are in the end infinitesimal neighbors on the macroscopic scale and nodes which are lying some distance apart on the coarse-grained scale can not easily be distinguished on the primordial scale. The near/far-order on the macroscopic scale is a result of the renormalization algorithm and is largely hidden on the microscopic scale. (As to the technical steps see \cite{Requ1},\cite{Requ2},\cite{Requ8}).

What we have just said is however only the geometric or space-time part of the task. It is important that we do not completely wipe out the translocal interaction structure in the coarse-graining steps. On the other hand, it is evidently not easy to express or envisage this translocal features within a smooth continuous description of space-time. Note that the microscopic model system has more DoF and microscopic relations among them as can be represented on the level of a continuous space-time with its built in near order structure. That is, we must find a way to represent at least certain collective aspects of this translocal structure within the restricted possibilities of a continuous space-time structure. This problem is closely related to the problem of \tit{polydimensions} we  mentioned in the introductory section of this paper. 

 We called such a structure a \tit{wormhole space} (cf. \cite{Requ3},\cite{Loch}). The above Kuramoto model, which inherits its near/far-order via the (quasi-Euclidean) lattice metric given on $\Z^2$ is a simple example. In a continuum limit we have to keep track of the sparse network of translocal connections. In our much more complicated networks there is no natural metric which survives automatically the renormalization steps. The networks themselves are dynamic and erratic and only certain collective DoF survive the coarse-graining steps. Note that the emergence of macroscopic observables is already a problem in ordinary quantum theory (cf. the discussion in \cite{deco}).   

In \cite{Requ3} section 4.1,4.2 we managed to derive formulas and make instructive numerical estimates of the number of, for example, microscopic wormholes connecting different regions of macroscopic space-time. On atomic scales we got:
\begin{ob} Whereas the network of translocal connections turned out to be quite sparse on the Planck scale the number of microscopic wormholes is nevertheless substantial on atomic scales. Take e.g. two regions of diameter $l_a$  being of the order of $10^{-10}$m and being a distance of $1$m apart. The number of translocal connections, $N_{tr}$ is in 3-dimensional space of the order $10^{79}$. Even if we take the diameter of our universe, i.e. $R_0\approx10^{10}$ly, we have still $N_{tr}\approx 10^{26}$. The reason for that behavior are the  scaling properties of these formulas derived in \cite{Requ3} which are a consequence of the assumed holographic principle.
\end{ob}
Furthermore, due to the sparseness of the translocal network, these quantities behave approximately additive. That is, The number of connections to a volume $V=\cup_iV_i$ with $V_i$ disjoint sets is roughly the sum of the connections to the volumes $V_i$. This is the consequence of the following observation.
\begin{ob}In sparse (random) graphs the property of spreading holds. I.e., certain graph properties are approximately additive (cf. sect. 4.1 of \cite{Requ3} and \cite{Bollo} p.255). 
\end{ob}

On the level of the primordial network, either in the form of nodes and edges (i.e., considered as a graph, $G$,) or elementary oscillators and their interactions, we can make some useful definitions. If a node, $x_i$, in a graph $G$ is connected to a node $x_j$ by an edge (depicted as $(x_i,x_j)$), we denote this by $x_i\sim x_j$. Given two subgraphs, $G_1,G_2$ of a graph $G$  
 with node sets $\{x^1_i\},\{x^2_j\}$, we can define the set of interbonds $(x^1_i,x^2_j)$, connecting nodes in $G_1$ with nodes in $G_2$. For simplicity reasons we assume that their node sets are disjunct. We can then compare the cardinality of this set, $|G_1\sim G_2|$, with the cardinality of the maximal possible set, $|G_1\sim G_2|_{max}=|G_1|\cdot|G_2|$ ($|G_i|$ the number of nodes in $G_i$).
\begin{defi}We define the connectivity of the disjunct subgraphs, $G_1$ and $G_2$, by
\beq c_{G_1G_2}:=  |G_1\sim G_2|/|G_1\sim G_2|_{max}   \eeq
\end{defi}
\begin{ob}In the random graph framework $c_{G_1G_2}$ denotes the probability of the existence of an edge between two randomly selected nodes in $G_1,G_2$ respectively.
\end{ob}
\begin{bem}This concept can in particular be applied to two cliques in the graph $G$ which roughly decribe infinitesimal neighborhoods of 'physical points'.
\end{bem}
The above described microscopic wormholes are such interbonds, connecting the macroscopic regions, $V_1,V_2$, say. Our above results about the distribution of such wormholes show that the connectivity between two possibly widely separated regions of space is in fact very low but non-vanishing.

We will calculate this connectivity for the above example of two volume elements of diameter $10^{-10}$m being 1m apart. In a volume element of this size there exist roughly $10^{75}$ grains of Planck-size. We thus have
\begin{conclusion}These two volume elements, or, rather, the two subgraphs, corresponding to them on the Planck level, have a connectivity
\beq c_{G_1G_2}=10^{79}/(10^{75})^2=10^{-71}   \eeq
Expressed in probabilities, the probability that two arbitrary nodes in the two volume elements are connected by an edge or two planck-size grains are connected by a microscopic wormhole is $10^{-71}$.
\end{conclusion}

In the following we want to concentrate, for the time being, on the simplest scenario, i.e., classical (uncurved) space-time as background structure plus non-relativistic quantum theory. This has the advantage of representing a relatively universal framework. Relativistic effects and, for example, general relativity shall be treated in future work. We remarked already above that the very existence of a smooth continuous space-time is in our view a consequence of the existence of a translocal deep structure being impressed upon our microscopic oscillator network. One should note that the existence of such a \tit{super structure} entails a vast shrinking of the microscopically available phase space which is quite typical for such complex entangled networks. 
\begin{conjecture} 
As a consequence we will assume that our network will essentially evolve on a (possibly complicated) attracting set which can be qualitatively described by a relatively small number of macroscopic (or, rather, mesoscopic) collective parameters.
\end{conjecture}

It is well known that such networks display a certain propensity to \tit{self-organization} as it is discussed in the context of \tit{synergetics}. We discussed these aspects briefly in \cite{Requ7} where more relevant literature is mentioned. 
\begin{defi}In the recent \cite{Grav} we called such a smooth space-time structure an order parameter manifold.
\end{defi}
\begin{ob}A characteristic property of the mentioned attracting set is the existence of a classical smooth metric structure and, in particular, the existence of a universal time concept. On a more fundamental scale this is given by the existence of an oscillation parameter (introduced previously)
\beq \omega_0:=\omega_{Pl}  \eeq
describing the ground oscillation of our oscillator network.
\end{ob}
\begin{conjecture}We conjecture that all our different time concepts we are employing on more macroscopic scales (e.g. cosmological time, thermodynamic time, etc.) derive ultimately from the existence of this ground oscillation in the deep structure of the quantum vacuum and the time scale it defines.
\end{conjecture}
\begin{bem}In the coarse graining we averaged out the erratic short scale oscillations, $\delta A_i(t),\varphi_i(t)$ on the more primordial scales.
\end{bem}

In a next step we have to distill the other relevant collective parameters which describe the attracting set we experience macroscopically as our particular space-time and the quantum excitations living in it. They are of course related to the conjectured   
large scale, slowly varying excitations $(a_i(t),\theta_i(t))$ we introduced previously. In the following, in order to fully understand the emergence of \tit{polydimensions}, we divide our analysis into the quantum mechanical one-particle case and the case of several particles (particles understood in the sense of traditional quantum theory).

It is noteworthy that Schroedinger in his series of fundamental papers realized the necessity of \tit{complex structure} on an ontological level only rather lately. This complex structure was also a point which troubled Ehrenfest in his discussion with Pauli (mentioned in the introduction). Personally we think that in the complex structure of quantum mechanics the interaction and entanglement of two real quantities is encoded.
\begin{conjecture}In the complex structure of quantum mechanics the entanglement of two real quantities is encoded. The complex superposition principle reflects the coevolution of these two quantities.
\end{conjecture}

We follow Bohm by splitting the wave function into \tit{phase function} $S(x_1,\ldots,x_n;t)$ and $R:=(\bar{\psi}\psi)^{1/2}(x_1,\ldots,x_n;t)$. That is, in the one-particle case we have $\psi(x;t)=R(x;t)e^{iS(x;t)}$ and
\beq \partial_t(R^2)=-\nabla(R^2\cdot\mbf{v})\quad\text{with}\quad \mbf{v}:=m^{-1}\cdot\nabla S   \eeq
\beq \partial_t S=-(2m)^{-1}(\nabla S)^2+V-(\hbar^2/2m)\cdot\triangle R/R   \eeq
\begin{defi}The expression
\beq V_q:= -(\hbar^2/2m)\cdot\triangle R/R   \eeq
is called quantum potential by Bohm.
\end{defi}

For two quantum particles we have for example
\beq \partial_t(R^2)=-(\nabla_1(R^2\nabla_1 S/m_1)+\nabla_2(R^2\nabla_2 S/m_2))   \eeq
\begin{multline} \partial_t S=-((2m_1)^{-1}(\nabla_1 S)^2+(2m_2)^{-1}(\nabla_2 S)^2)+V(x_1,x_2)\\-((\hbar^2/2m_1)\triangle_1 R/R+(\hbar^2/2m_2)\triangle_2 R/R)   \end{multline}
\begin{bem}In most hidden variable theories these equations are interpreted within a classical or stochastic particle framework. In contrast to that point of view, our approach will be an undulatory one.
\end{bem}
\begin{ob}One should note that many people were intrigued by the similarity of the evolution equation of the phase function, $S$, to the classical \tit{Hamilton-Jacobi equation}. 
\end{ob}
\begin{bem}As an answer to Einsteins justified criticism of the linearity of quantum mechanics we see that the above coupled equations are non-linear. This also shows that the complex superposition principle is a highly non-trivial phenomenon (in contrast to its naive but actually unjustified plausibility).
\end{bem}
 
We begin with the one-particle case , given by a particular excitation pattern of the underlying network of oscillators of a certain characteristic complexity. We assume that our coarse-graining process leads finally on the level of ordinary quantum mechanics to the following excitation pattern.
\begin{conjecture}The one-particle case corresponds to a coarse-grained excitation pattern of the form
\beq q(x;t)=(A_0+a(x;t))\cdot\sin(\omega_0t+\theta(x;t))   \eeq
with the individual terms having the properties described above.
\end{conjecture}
We deleted in this representation the short-scale fluctuations $\delta A(x;t)$ and $\varphi(x;t)$. We hence assume that a one-particle excitation is represented on the attracting set in phase space by the collective quantities
\beq A_o\;,\; a(x;t)\; ,\; \omega_0t\; ,\; \theta(x;t)   \eeq
 \begin{bem} $\delta A(x;t),\varphi(x;t)$ do correspond to vacuum fluctuations. 
\end{bem}

In units where $\hbar=1$ we now make the following association:
\begin{assumption}We assume that $S(x;t)$ corresponds to $\theta(x;t)$ and $R(x;t)$ to $a(x;t)$, where in the latter case we assume that $a(x;t)$ is either positive or negative. We make the assumption that in the one-particle case an $a(x;t)$ that can be both positive and negative does not occur.
\end{assumption}
\begin{ob}The possibility of a positive or negative $a(x;t)$ allows to introduce particles and antiparticles. We regard a negative $a(x;t)$ as a hole (in the sense of Dirac) in the sea of vacuum fluctuations.
\end{ob}

As we are at the moment unable to solve the presumably quite complicated and, a fortiori, not really known microscopic evolution laws of the network, this has the status of a conjecture. We developed however a similar picture in a more accessible field, i.e., the study of \tit{thermal states} and their \tit{quasiparticle} excitation patterns (cf. e.g. section 3 of \cite{Hole1} and section 7 of \cite{Hole2}. We proved there that the Dirac picture of particle- and hole-excitations in the regime of thermal states makes sense. That is, we may conclude:
\begin{ob}[Dirac picture]The correspondence to our observations in the above cited papers makes it plausible that our assumptions, made above, are perhaps not so far-fetched.
\end{ob}

\begin{ob} In the above correspondence, i.e., action being equal to number of oscillations, 
\beq \omega_0\cdot t:=S_0(t)   \eeq
counts the change of phase of the ground oscillation (with respect to some arbitrary reference point/time) while $S(x;t)$ represents the space-time dependent phase modulation of the ground wave. $a(x;t)$, on the other hand, is an amplitude modulation.
\end{ob}
\begin{bem}The physical arbitrariness of the choice of reference point/time explains the global gauge invariance of the phase function in quantum mechanics.
\end{bem}
\begin{conclusion}The one-particle Schroedinger equation is in this philosophy of collective excitations and effective theories (cf. Bohm's remarks in the introduction) the representation of the coevolution of the two collective quantities
\beq a(x;t)=R(x;t)\quad\text{and}\quad \theta(x;t)=S(x;t)     \eeq
\end{conclusion}  

The case of several quantum particles represents a more complex type of excitation pattern and is more subtle. This is the place where the problem of \tit{polydimensions} comes into the play. The Schroedinger equation for several quantum particles with the above splitting has an $S$ and $R$ which depend on several positions at the same time, i.e. an $x_i$ for each particle $(i)$ (in the special case where all the particles can be distinguished by an observer). In our undulatory philosophy
\beq S(x_1,\ldots,x_n;t)\quad ,\quad R(x_1,\ldots,x_n;t)  \eeq
have to be associated with a certain excitation pattern of the oscillator network which is necessarily more complex and \tit{non-local} compared to the one-particle case. It is evident that the n-particle excitation must behave both differently in a characteristic way and configurationally stable so as to be distinguishable compared to the simpler one-particle situation.

In the one-particle case the coarse-graining was performed in the usual way by averaging over finer details on the next lower level in the neighborhood of the coarser variable, say, $x$. I.e., DoF, $q(x_i;t)$, with $x_i=x+\delta x_i$, $\delta x_i$ ranging in some neighborhood of $x=0$, are combined to a new variable
\beq q(x;t):=N^{-\alpha}\cdot\sum_{i=1}^N q(x_i;t)  \eeq
($\alpha$ a possible renormalization exponent).  

This simple procedure will not work in the multiparticle  case and shows the necessary non-triviality of the renormalization process under discussion. We have in fact to search for other and more specific \tit{relevant} collective DoF which have to be taken into account on the different levels of the process.
\begin{ob}As both the one-particle and the multiparticle excitations are in the end excitation patterns of the primordial oscillator network, they must be discernible on the surface level, i.e., within a framework which consists of classical smooth space-time plus ordinary quantum theory, by a peculiar collective excitation pattern which survives this renormalization process and which is different from the one-particle case.
\end{ob}
\begin{bem}One should note that the underlying oscillator network is expected to be capable of supporting quite a number of different collective excitation modes(think, e.g. of the normal modes in the comparatively simple example of crystal dynamics) in which, a fortiori, regions of space, which are a distant apart in classical space-time, may directly interact with each other.
\end{bem}   

We come back to our discussion of translocal connections at the beginning of this section. We observed that, while they are not so numerous as the connection to the immediate neighborhood of the (physical) points of our space-time manifold, there does nevertheless exist a sufficiently large number so as to make a contribution to the dynamics if they act \tit{coherently}. 
\begin{conjecture}We assume that these translocal interbonds (which, actually, represent elementary interactions) have a propensity to act coherently in case they are appropriately excited. That is, we assume that the collective observables
\beq S(x_1,\ldots ,x_n;t)\quad ,\quad R(x_1,\ldots ,x_n;t)  \eeq
represent a n-point collective excitation mode among the DoF in the respective infinitesmal neighborhoods of the points $x_i$.
\end{conjecture}
\begin{bem}In \cite{Requ7} we gave examples of microscopic models which describe such a translocal oscillation of certain charge carriers or carriers of information.
\end{bem}

That is, instead of a local phase- and amplitude modulation around the points of the space-time manifold in the one-particle case, we have now a translocal phase- and amplitude modulation of the underlying ground oscillation which includes n points at a time. We think this explanation gives the speculations of Schroedinger about individual wave patterns interpenetrating each other a concrete form.
\begin{ob}[Entanglement, EPR]This reasoning gives at the same time the notion of entanglement a concrete ontological meaning. Entangled quantum objects or regions of space-time are coupled to each other translocally via such collectively acting translocal bonds. By the same token, this holds for EPR-like experiments.
\end{ob}

We want to conclude this section with a remark concerning the problem of \tit{polydimensions}. In our view, both Schroedinger and Heisenberg are partly right. One can explain the multiparticle case within an undulatory framework but only at the expense of giving up locality. 
\begin{conclusion}[Polydimensions]Schroedinger's undulatory picture can be maintained at the expense of having a more complicated model of space-time. As far as we can see, this latter possibility was neither entertained by Schroedinger nor Heisenberg. But, in our view, it is the decisive element of quantum theory.
\end{conclusion} 
\section{The Collapse of the Wave Packet and the Hidden Non-Locality of Quantum Theory}
While in N-particle quantum theory a possible non-locality is more apparent, it is somewhat hidden in the formulation of one-particle quantum mechanics. The same holds for the quantum measurement process in general. The Schroedinger equation is strictly local in its appearance and we want to show in a first step that this is entirely compatible with an underlying hidden translocal behavior. 

To illustrate this point we discuss two (consecutive) stages in our renormalization process with the final level (II) being the level of ordinary quantum theory, the next lower level (I) (or perhaps some other more microscopic stage) displaying a certain non-locality. We assume that the evolution equations on level (II) are of a local character, to give an example:
\beq \dot{U}(x;t)=\triangle V(x;t)+\ldots   \eeq
for, say, two quantities, $U(x;t),V(x;t)$. 

On the next lower level we may instead have microscopic quantities, $u(x;t),v(x;t)$, with their respective non-local evolution equations. Furthermore, we may have the relation (between the two levels):
\beq V(x;t)=\int F(x,y;t)\cdot v(y;t)dy\quad ,\quad U(x;t)=\int G(x,y;t)\cdot u(y;t)dy   \eeq
with $F,G$ being some non-local \tit{influence functions} (note that these are very simple examples; the situation can of course be much more involved).
\begin{bem}A similar scenario we discussed in our analysis of holography and the bulk-boundary dependence in \cite{Requ3}.
\end{bem}

This then yields
\beq \dot{U}(x;t)=\int \triangle_x F(x,y;t)\cdot v(y;t)dy+\ldots    \eeq
\begin{conclusion}The two consecutive renormalization levels connect two effective theories, the one with a possible non-local evolution equation, the other with a local description while the collective variables of the upper level depend non-locally on the collective, but more microscopic variables of the next lower level.
\end{conclusion}  
In non-relativistic quantum theory the role of $U,V$ is played by $S$ and $R$. In \cite{Requ7} we already formulated such non-local laws, connecting $R,S$ with microscopic DoF, living on the primordial network level. At the time of writing \cite{Requ7} we however did not have the estimates about translocal connections at our disposal which followed from our analysis of the \tit{holographic principle} made in \cite{Requ3}.

In one-particle quantum theory the non-locality manifests itself only through measurement interferences or similar operations. We assume that a quantum operation, induced by the measurement of some observable takes place in a localized region of space-time. After all it is the interaction of a certain macrosystem, being in a particular many-body state, with the wave function of the quantum object. The information of the interaction then travels via the existing translocal connections to every other region of configuration space.
\begin{conclusion}The existence of the translocal connections, hidden in ordinary classical space-time, spreads the measurement interference to all points of space, thus making possible a (quasi) instantaneous reduction to a different wave-function, for example, a sharply localized one, which we interpret in the conventional measurement scheme as the detection of a particle. Put differently,  the slowly varying amplitude- and phase modulation is changed globally or even wiped out instantaneously in certain regions of space.
\end{conclusion}

As a last point we want to discuss the non-commutativity of observables as a consequence of translocality. 
\begin{ob}As the spectral theorem tells us that observables can essentially be reduced to projections, it suffices to discuss measurement operations. A measurement of, say, observable $A$ at time $t$ changes instantaneously the wavefunction, $\psi(t)$, (as long as it is not an eigenfunction of $A$) to some $\psi'(t+\varepsilon)$. A consecutive measurement of an observable, $B$, at time $(t+\varepsilon)$ acts on the altered wavefunction  $\psi'(t+\varepsilon)$ changing it again in general. If we exchange $A$ and $B$ we get in general a different result, i.e.:
\beq P_A P_B\psi\neq P_B\ P_A\psi  \eeq
as a consequence of the translocality of our theory.
\end{ob}

We have so far only sketched some of the quantum phenomena which can be adressed within our framework. A more detailed analysis of possible consequences shall be given in future work.   


\begin{thebibliography}{99} {\small
 \bibitem{Moore}W.Moore: ``Schroedinger, Life and Thought'', Cambridge Univ.Pr., Cambridge 1989
\bibitem{Wheeler}C.W.Misner,K.S.Thorne,J.A.Wheeler: ``Gravitation'', Freeman, San Francisco 1970
\bibitem{Adler}S.Adler: ``Quantum Theory as an Emergent Phenomenon'', CambridgeUniv.Pr., Cambridge 2004
\bibitem{Pad}T.Padmanabhan: ``Lessons from Classical Gravity about the Quantum Structure of Spacetime'', arXiv:1012.4476
\bibitem{Verlinde}E.Verlinde: ``On the Origin of Gravity and the Laws of Newton'', JHEP 1104(2011)029, arXiv:1001.0785
\bibitem{Vienna}Conference: ``Emergent Quantum Mechanics'', Vienna, November 2011
\bibitem{Bohm}D.Bohm: ``Wholeness and the Implicate Order'', Routhledge and Kegan, London 1980
 \bibitem{Weinberg}S.Weinberg: ``Dreams of a final Theory'', Vintage, London 1993
 \bibitem{Wald}R.M.Wald: ``Quantum Field Theory in Curved Spaetime'', Univ. of Chicago Pr., Chicago 1994
 \bibitem{Cao}T.Y.Cao: ``Conceptual Developments of 20th Century Field Theories'', Cambridge Univ. Pr., Cambridge 1997
 \bibitem{Dirac}P.A.M.Dirac: ``Relativity and Quantum Mechanics'', Fields and Quanta 3(1972)139, Gordon and Breach, London 1972
 \bibitem{Yang}C.N.Yang:
``Square root of minus one, complex phases and E.Schroedinger'', Schroedinger, Centenary of a
Polymath, Ed. C.W.Kilmister, Cambridge Univ.Pr., London 1987
 \bibitem{Ehrenfest}P.Ehrenfest: ``Einige die Quantenmechanik betreffende Erkundigungsfragen'', Z.Phys. 78(1932)555
\bibitem{Pauli1}W.Pauli: ``Einige die Quantenmechanik betreffende Erkundigungsfragen'', Z.Phys.
80(1933)573
 \bibitem{Pauli2}W.Pauli: ``Wiss. Briefw.'' Vol II, Ed. K.v.Meyenn, Springer, Berlin 1985
\bibitem{Sync}S.Strogatz: ``Sync'', Penguin Pr., London 2003
 \bibitem{Watts}D.J.Watts: ``Small Worlds``, Princeton Univ.Pr., Princeton 1999
 \bibitem{Requ1}M.Requardt: ``A geometric renormalization group in discrete quantum space-time'', J.Math.Phys. 44(2003)5588, arXiv: gr-qc/0110077
 \bibitem{Requ2}M.Requardt: ``(Quantum) spacetime as a statistical geometry of lumps in
random networks'', Class.Quant.Grav. 17(2000)2029
 \bibitem{Pines}D.Bohm,D.Pines: Phys.Rev. 85(1953)338, 92(1953)609
 \bibitem{Requ3}M.Requardt: ``Wormhole Spaces: the common Cause for the
Black Hole Entropy Area Law, the Holographic Principle and Quantum Entanglement'', arXiv:0910.4017
\bibitem{Requ4}M.Requardt: ``Planck Fluctuations, Measurement Uncerttainties and the Holograhic
Principle'', Mod.Phys.Lett. A 22(2007)791, arXiv:gr-qc/0505019
 \bibitem{Requ5}M.Requardt: ``About
the Minimal Resolution of Space-Time Grains in Experimental Quantum Gravity'', arXiv:0807.3619
\bibitem{Rugh1}S.E.Rugh,H.Zinkernagel,T.Y.Cao: ``The Casimir Effect and the Interpretation of the
Vacuum'', Stud.Hist.Phil.Mod.Phys. 30(1999)111
 \bibitem{Rugh2}S.E.Rugh,H.Zinkernagel: ``The Quantum
Vacuum and the Cosmological Constant Problem'', Stud.Hist.Phil.Mod.Phys. 33(2002)663
\bibitem{Boyer}T.H.Boyer: ``Quantum Zero-Point Energy and Long-Range Forces'', Ann.Phys. 56(1970)474
\bibitem{Weinberg2}S.Weinberg: ``The Cosmological Constant Problem'', Rev.Mod.Phys. 61(1989)1
\bibitem{Strau}N.Straumann: ``On the Mystery of the Cosmic Vacuum Energy Density'', Eur.J.Phys.
20(1999)419, arXiv:astro-ph/0009386
\bibitem{Wheeler2}J.A.Wheeler: ``Superspace and the Nature of Quantum Geometric Dynamics'', Battelle Rencontres 1967, eds. B.DeWitt, J.A.Wheeler, Benjamin, N.Y.1968
 \bibitem{Requ6}M,Requardt: ``Fluctuation Operators and
Spontaneous Symmetry Breaking '', J.Math.Phys. 43(2002)351, arXiv:math-ph/0003012
\bibitem{Witt}A.Anderson,B.DeWitt: ``Does the Topology of Space fluctuate?'', Found.Phys. 16(1986)91
\bibitem{Hooft1}G.'t Hooft: ``How doe God play Dice?,(Pre)Determiism at the Planck Scale'', an essay
in honour of J.S.Bell, arXiv:hep-th/0104219
 \bibitem{Hooft2}G.'t Hooft: ``Quantum Gravity as a
Dissipative Determinstic System'', CQG 16(1999)3263, arXiv:gr-qc/9903084
\bibitem{Now}T.Nowotny,M.Requardt: ``Emergent Properties in Structurally Dynamic Cellular
Automata'', J.Cell.Aut. 2(2007)273, arXiv:cond-mat/0611427
 \bibitem{Loch}A.Lochmann,M.Requardt: ``An Analysis of the Transition Zone between the various Scaling Regimes in the Small World Model'', J.Stat.Phys. 122(2006)255, arXiv:cond-mat/0409710
\bibitem{Requ8}M.Requardt: ``The Continuum Limit of Discrete Geometries'', Int.J.Geom.Meth.Mod.Phys. 3(2006)285, arXiv:math-ph/0507017
\bibitem{Stro1}S.H.Strogatz: ``From Kuramoto to Crawford: exploring the onset of synchronization in populations of coupled oscillators'', Physica D 143(2000)1
\bibitem{Stro2}S.H.Strogatz: ``Nonlinear Dynamics and Chaos'', Perseus Books, Cambridge MA 2000
\bibitem{Jackson}E.A.Jackson: ``Properties of nonlinear dynamics I'', Cambridge Unv.Pr., Cambridge 1990
\bibitem{Nie}E.Niebur,H.G.Schuster,D.M.Kammen,C.Koch: ``Oscillator-phase coupling for different two-dimensional network connectivities'', P.R. A 44(1991)6895
\bibitem{Requ7}M.Requardt: ``Let's call it Nonlocal Quantum Physics'', arXiv:gr-qc/0006063
\bibitem{Bog}N.N.Bogoljubov,J.A.Mitropolski: ``Asymptotische Methode in der Theorie der nichtlinearen Schwingungen'' 3.ed., Akademie-Verlag, Berlin 1965
\bibitem{Sanders}J.A.Sanders,F.Verhulst,J.Murdock: ``Averaging Methods in Nonlinear Dynamical Systems'' 2.ed., Springer, Berlin 2007 
\bibitem{deco}M.Requardt: ``An Alternative to Decoherence by Environment and the Appearance of a Classical World'', arXiv:1009.1220
\bibitem{Bollo}B.Bollobas: ``Random Graphs'', sec.ed., Cambridge Univ. Pr., Cambridge 2001
\bibitem{Grav}M.Requardt: ``Gravitons as Goldstone Modes and the Spontaneous Symmetry Breaking of Diffeomorphism Invariance'', arXiv:1203.1702
\bibitem{Hole1}M.Requardt: ``A structure theorem of general KMS-states with a possible bearing on the construction of creation and annihilation operators for collective excitations and holes'', J.Phys.A: Math.Gen. 18(1985)287
\bibitem{Hole2}M.Requardt: ``Spontaneous Symmetry Breaking of Lorentz and Galilei Boosts in (Relativistic) Many-Body Systems'', arxiv:0805.3022




} \end{thebibliography}
 \end{document}